\documentclass[%
 reprint,
 superscriptaddress,
 showpacs,
 showkeys,
 amsmath,amssymb,
 aps,
 prd,
]{revtex4-1}

\usepackage{graphicx}
\usepackage{dcolumn}
\usepackage{bm}
\usepackage{hyperref}

\begin{document}


\title{$K_{0}^{\ast}(800)$ as a companion pole of $K_{0}^{\ast}(1430)$}

\author{Thomas \surname{Wolkanowski}}
 \email{wolkanowski@th.physik.uni-frankfurt.de}
\affiliation{%
 Institute for Theoretical Physics, Goethe University, D-60438 Frankfurt am Main, Germany}%


\author{Milena \surname{So\l tysiak}}
 \email{milena.soltysiak@op.pl}
\affiliation{%
 Institute of Physics, Jan Kochanowski University, PL-25406 Kielce, Poland}%

\author{Francesco \surname{Giacosa}}
 \email{fgiacosa@ujk.edu.pl}
\affiliation{%
 Institute for Theoretical Physics, Goethe University, D-60438 Frankfurt am Main, Germany}%
\affiliation{%
 Institute of Physics, Jan Kochanowski University, PL-25406 Kielce, Poland}%


\date{\today}

\begin{abstract}
We study the light scalar sector up to $1.8$ GeV by using a quantum field theoretical approach which includes a single
kaonic state in a Lagrangian with both derivative and non-derivative interactions. By performing a fit to $\pi K$ phase
shift data in the $I=1/2,$ $J=0$ channel, we show that $K_{0}^{\ast}(800)$ (or
$\kappa$) emerges as a dynamically generated companion pole of $K_{0}^{\ast
}(1430)$. This is a result of investigating quantum fluctuations with one kaon
and one pion circulating in the loops dressing $K_{0}^{\ast}(1430)$. We determine the position of the poles on the complex plane in the context of our approach: 
for $K_{0}^{\ast}(1430)$ we get $(1.413\pm0.002)-i\hspace{0.02cm}(0.127\pm0.003)$ (in GeV),
while for $\kappa$ we get $(0.746\pm0.019)-i\hspace{0.02cm}(0.262\pm0.014)$
(in GeV). The model-dependence of these results and related uncertainties are discussed in the paper. 
A large-$N_{c}$ study confirms that $K_{0}^{\ast}(1430)$ is
predominantly a quarkonium and that $K_{0}^{\ast}(800)$ is a molecular-like
dynamically generated state.
\end{abstract}

\pacs{12.40.Yx, 13.75.Lb, 13.30.Eg, 11.55.Fv}
\keywords{scalar kaons, companion pole, dynamical generation}
\maketitle


\section{\label{sec:introduction}Introduction}
The lightest scalar resonance with isospin $I=1/2$ is the state $K_{0}^{\ast
}(800)$, also denoted as $\kappa$. This state is not yet listed in the summary
table of the Particle Data Group (PDG) \cite{olive}. The confirmation of
$\kappa$ is important, since it would complete the nonet of light scalar
states below $1$ GeV. Namely, besides the putative $\kappa$ state, the broad
but by now established $f_{0}(500)$ (see Ref.\ \cite{sigmareview} and
references therein) as well as the narrow resonances $a_{0}(980)$ and
$f_{0}(980)$ are well-established mesons \cite{olive}. These light scalar
mesons are excellent candidates to be non-conventional states, $i.e.$,
four-quark objects, realized as diquark-antidiquark states
\cite{jaffe,*jaffe2,jaffe3,maiani,tqmix2,*tqmix,*pagliaraderivatives2,fariborz2,*fariborz,*rodriguez,*fariborz3}
and/or as dynamically generated molecular-like states
\cite{tornclose,pelaez,*pelaez2,pelaez3,oller,*oller2,*oller3,oller4,oller5,*oller6,2006beveren,
morgan,dullemond,tornqvist,*tornqvist2,boglione,*pennington}
(for review, see also Ref.\ \cite{amslerrev,*amslerrev2}).

The aim of this work is to apply a quantum field theoretical approach in order
to investigate the existence of the $\kappa$ as well as its nature. Within our
approach a \textit{single} (quark-antiquark) seed state, roughly corresponding
to the well-known resonance $K_{0}^{\ast}(1430)$, is described by an effective
Lagrangian. In particular, we shall use a Lagrangian that contains -- in agreement with
chiral perturbation theory (chPT) and chiral models -- both derivative and
non-derivative interaction terms. As we shall see, the simultaneous presence
of both of them ensures a good description of scattering data. Indeed, as
expected from chPT the derivative interaction gives the largest contribution.
 After computing the full one-loop resummed propagator we perform a
fit to experimental $\pi K$ phase shift data from Ref.\ \cite{pionkaonexp}. The
fit depends on four parameters of the model: two coupling constants, one bare
mass, and one cutoff entering a Gaussian form factor. We find that, besides the expected resonance pole of
$K_{0}^{\ast}(1430)$, a pole corresponding to the light $\kappa$ naturally
emerges on the unphysical Riemann sheet. In this situation the $\kappa$ is
established as a dynamically generated companion pole of the conventional
quark-antiquark meson $K_{0}^{\ast}(1430)$. We determine the position of the
poles for both states including errors. For previous determinations of the
pole position of $\kappa$ see $e.g.$ Refs.\
\cite{ishida2,magalhaes,descotes,pelaez2,pelaez3,zheng,*zheng2,black,fariborzk,ledwig}, as
well as the experimental observation by BES \cite{beskappa} and the lattice
study of Ref.\ \cite{fu}.

Moreover, $(i)$ it turns out that the light $\kappa$ does not correspond to
any peak in the scalar kaonic spectral function but only to an enhancement in
the low-energy region at about $750$ MeV. A large-$N_{c}$ study shows that its
pole disappears when $N_{c}$ is large enough ($N_{c}\simeq13$). As a
consequence, this state is \textit{not} predominantly a quarkonium but rather
a dynamically generated meson. $(ii)$ On the other hand, the pole of the
corresponding state above $1$ GeV tends to the real energy axis in the
large-$N_{c}$, as expected for a predominantly quark-antiquark state.

For completeness, we also investigate the statistical significance of our
results: we find that both derivative and non-derivative interactions are
needed for a satisfactory fit. On the contrary, variations of the models with
only derivative or non-derivative interactions or with other form factors
different from the Gaussian turn out not to be in agreement with the
experimental results.

\section{\label{sec:model}The model}
Our model consists of an interaction Lagrangian describing the
interaction/decay of a single scalar kaonic seed state, denoted as
$K_{0}^{\ast}$, into one pion and one kaon. In agreement with effective
approaches of low-energy QCD (both chPT \cite{chpt,*Ecker,*chpt2} and
effective chiral models \cite{ko,*ko2,elsm2}, based on the nonlinear and
linear realization of chiral symmetry, respectively), it consists of two types
of terms, $i.e.$, one without and one involving derivatives:
\begin{eqnarray}
\mathcal{L}_{\text{int}} & = & aK_{0}^{\ast-}\pi^{0}K^{+}+bK_{0}^{\ast-}\partial_{\mu}\pi^{0}\partial^{\mu}K^{+} \label{eq:lag} \\
& & + \ \sqrt{2}aK_{0}^{\ast-}\pi^{+}K^{0}+\sqrt{2}bK_{0}^{\ast-}\partial_{\mu}
\pi^{+}\partial^{\mu}K^{0}+\dots \ , \nonumber
\end{eqnarray}
where dots represent analogous interaction terms for the other members of the
isospin multiplets, as well as Hermitian conjugation. The decay width as
function of the (running) mass $m$ of the unstable $K_{0}^{\ast}$ reads
\begin{equation}
\Gamma_{K_{0}^{\ast}}(m)=3\hspace{0.02cm}\frac{k(m)}{8\pi m^{2}}\left[
a-b\hspace{0.02cm}\frac{m^{2}-m_{\pi}^{2}-m_{K}^{2}}{2}\right]^{2}%
F_{\Lambda}(m)\ ,\label{eq:width}%
\end{equation}
where the factor of $3$ comes from summing over isospin. Here, we introduced
the modulus of the three-momentum of the outgoing particles in the rest frame
of the decaying particle as
\begin{align}
k(m)  & =\frac{\sqrt{m^{4}+\left(  m_{\pi}^{2}-m_{K}^{2}\right)  ^{2}-2\left(
m_{\pi}^{2}+m_{K}^{2}\right)  m^{2}}}{2m}\nonumber\\
& \times\ \theta\left(  m-m_{\pi}-m_{K}\right)  \ .
\end{align}
The quantities $m_{\pi}$ and $m_{K}$ are the pion and kaon mass, respectively.
The form factor $F_{\Lambda}(m)$ is chosen as
\begin{equation}
F_{\Lambda}(m)=e^{-2k^{2}(m)/\Lambda^{2}}\ ,\label{eq:formfactor}%
\end{equation}
where $\Lambda$ is an energy scale arising from the fact that mesons are not
elementary objects (technically, it can be included already in the Lagrangian
by making it non-local, see $e.g.$ Ref.\ \cite{nonlocal,*nonlocal2}). This
parameter acts as a cutoff and assures that all our calculations are finite.

When the form factor is set to zero in Eq.\ (\ref{eq:width}) and $m\simeq1.43$
GeV, we obtain the so-called tree-level decay width. It can be identified with
the physical width of the $K_{0}^{\ast}(1430)$ in (some) phenomenological
models, in which this resonance is interpreted as a quarkonium \cite{elsm2}.
As we shall see, the bare seed state $K_{0}^{\ast}$ in our Lagrangian\
(\ref{eq:lag}) in fact corresponds roughly to the well-known resonance
$K_{0}^{\ast}(1430)$ -- this is in agreement with various phenomenological
studies of the scalar sector
\cite{fariborz2,*fariborz,*rodriguez,*fariborz3,elsm2,close01,*close3,*close7,*close5,isgur}.
\begin{figure}[h]
\centering
\includegraphics[scale=0.27]{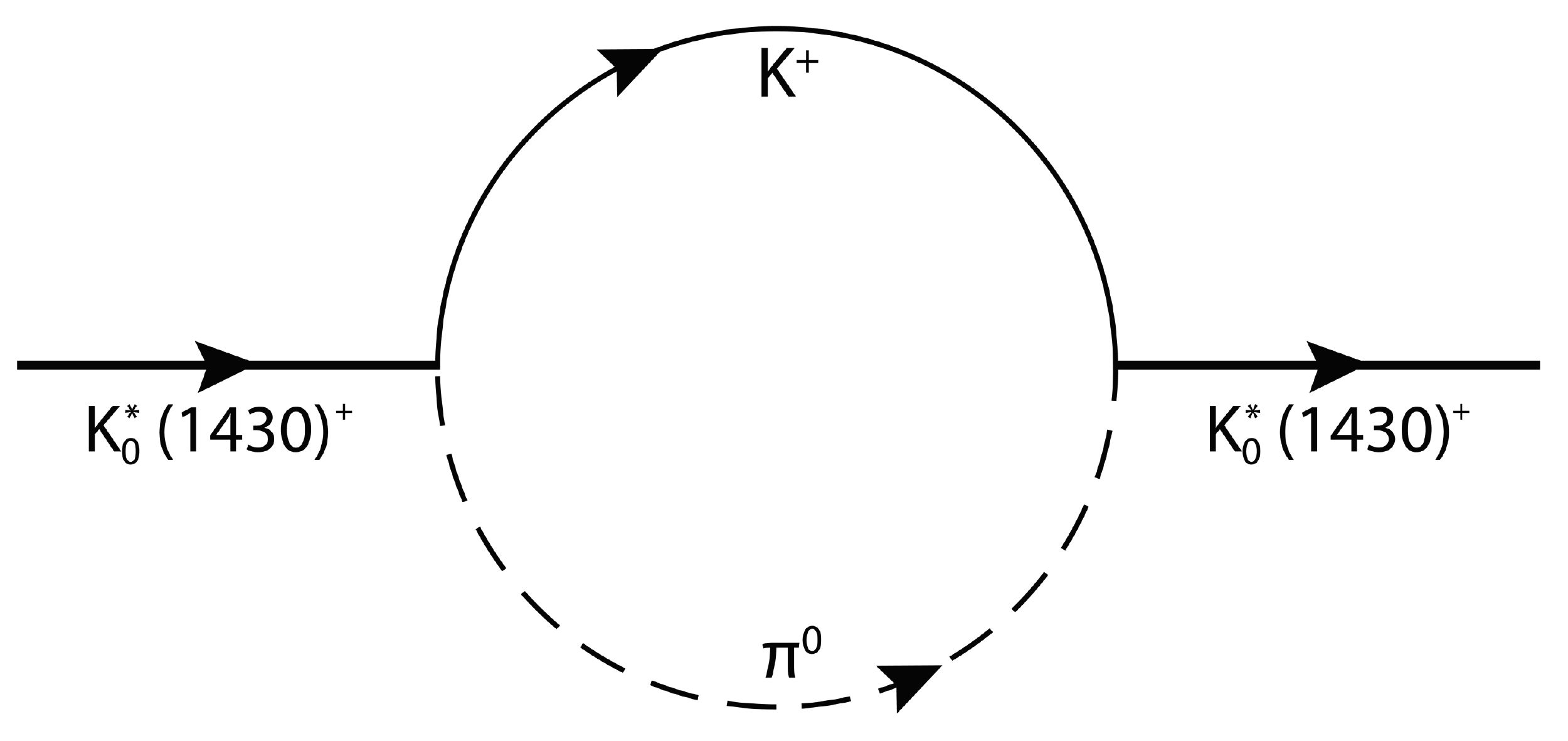}\caption{\label{fig:loop}Example of a one-loop
contribution to $\Pi(m)$.}%
\end{figure}

Following closely Ref.\ \cite{a0} (see also Refs.\
\cite{dullemond,tornqvist,*tornqvist2,boglione,*pennington,giacosapagliara,*e38}%
), we now briefly present the mathematical formalism. The propagator of the
scalar kaonic field is given by
\begin{equation}
\Delta_{K_{0}^{\ast}}(p^{2}=m^{2}) = \frac{1}{m^{2}-m_{0}^{2}-\Pi
(m)+i\varepsilon} \ ,
\label{eq:propagator}
\end{equation}
where $m_{0}$ is the bare mass of the scalar kaon and $\Pi(m)$ is the sum of
all one-loop contributions with one pion and one kaon circulating in it, see
Fig.\ \ref{fig:loop}. Although the loops in our model are regularized by the
form factor in Eq.\ (\ref{eq:formfactor}), one has to take into account
emerging tadpole diagrams when using ordinary Feynman rules. 
The details are discussed in Ref.\ \cite{a0}. A study of the
validity of the one-loop approximation was done in Ref.\ \cite{jonas}. We will
use that the spectral function is obtained from the propagator by
\begin{equation}
d_{K_{0}^{\ast}}(m) = -\frac{2m}{\pi}\operatorname{Im}\Delta_{K_{0}^{\ast}%
}(p^{2}=m^{2}) \ ,\label{eq:spectral}%
\end{equation}
having the correct normalization $\int_{0}^{\infty}d_{K_{0}^{\ast}%
}(m)\mathrm{dm}=1$, and that according to the optical theorem
$\operatorname{Im}\Pi(m)=-m\Gamma_{K_{0}^{\ast}}(m)$.

The $J=0$ and $I=1/2$ phase shift for $\pi K$ scattering up to $1.8$ GeV is
assumed to be dominated by the scalar kaonic resonances(s). Within our
framework it therefore takes the form (see the review of kinematics provided
by the PDG \cite{olive})
\begin{equation}
\delta_{\pi K}(m) = \frac{1}{2}\arccos\left[ 1-\pi\Gamma_{K_{0}^{\ast}%
}(m)d_{K_{0}^{\ast}}(m)\right]  \ .\label{eq:phaseshift}%
\end{equation}

Some comments are in order:\\
$(i)$ Eq.\ (\ref{eq:phaseshift}) is based on the assumption that the $s$-channel propagation
dominates, c.f.r. Ref.\ \cite{olive}. The validity of this assumption (and thus
neglecting the contributions from the $u$-channel exchange diagrams) was
extensively discussed in the literature\ \cite{isgurcomment,*tornreply1,harada,*harada2,ruppcomment}. 
In particular, it was shown that this approximation alters only slightly the position of the resonance
poles: it is therefore very suitable for our purposes.\\
\noindent$(ii)$ Furthermore, the approximation of keeping only the $s$-channel is
justified by the fact that we perform a fit to data starting at about $200$
MeV above the $\pi K$-threshold. This is far enough from the threshold, where
the overall interaction strength is small and all contributions are relevant
(and where chiral symmetry is especially important, see also the
considerations in the next point).\\
\noindent$(iii)$ Note also that we do not use any constant background term in our
model. This is different from many previous works on the subject (see $e.g.$
Ref.\ \cite{ishida2} or, more recently, Ref.\ \cite{fariborzk}); instead, we utilize derivative
interactions. In order to illustrate this point, we introduce an analogy with
the old linear sigma model, which contains a non-derivative interaction as
well as a back-ground term. The potential of the model has the usual Mexican
hat form, $V=\frac{\lambda}{4}(\vec{\pi}^{2}+\sigma^{2}-F^{2})^{2}%
-\varepsilon\sigma.$ The field $\sigma$ has a non-vanishing vacuum expectation
value $\phi;$ as a consequence (after performing the shift $\sigma
\rightarrow\sigma+\phi$) the mass of $\sigma$ reads $M_{\sigma}^{2}%
=\lambda\phi^{2}$, while the pion mass reads $M_{\pi}^{2}=\varepsilon/\phi$
and vanishes in the chiral limit (where $\varepsilon\propto m_{q}$ vanishes).

Retaining only the interaction terms relevant for $\pi\pi$ scattering, we have
$V=$ $\frac{\lambda}{4}\vec{\pi}^{4}+\lambda\sigma\vec{\pi}^{2}+\dots$, thus
one is left with a non-derivative interaction through $\sigma$-exchange, as
well as a four-leg repulsion term. After transforming the fields into a polar
form by $(\sigma,\vec{\pi})\rightarrow\sigma e^{i\vec{t}\cdot\vec{\pi}}$ (an
intermediate step toward chiral perturbation theory), we obtain $V=\frac
{1}{\phi}\sigma(\partial_{\mu}\vec{\pi})^{2}-\frac{M_{\pi}^{2}}{2\phi}%
\sigma\vec{\pi}^{2}+\dots$, $i.e.$, no background term of type $\vec{\pi}^{4}$
is present, but a dominant derivative interaction has emerged.

The non-derivative interaction is subdominant and vanishes in the chiral
limit: this is in agreement with low-energy chiral theorems. The interchange
of one pion field with one kaon field allows us to pass from the case of the
$\sigma$ to that of the kaonic sector studied here (formally, it is a simple
rotation in flavor space), but the very same intuitive arguments show why the
use of derivative interactions is important for scalar mesons in general.
Moreover, the contemporary presence of derivative and non-derivative
interactions implies that the structure giving rise to Adler's zero is
automatically fulfilled (we thus do not have to add the Adler's zero
separately, as done for example in Ref.\ \cite{zhiyong}).\\
\noindent$(iv)$ Our model is designed to study the scattering in the $I=1/2$ channel
only, in which the $s$-wave exchange of a scalar kaon can be considered as
dominant. Indeed, the scalar kaon contributes also through $u$-channel
exchange diagrams to the cross-section. Experimentally, the $I=3/2$ phase
shift is negative ($i.e.$, there is a repulsion in this channel) but is at
least a factor of $4$ smaller than for $I=1/2,$ showing also that the enhanced
intensity in the $I=1/2$ channel can be ascribed to the $s$-wave exchange of a
scalar kaon.

\section{\label{sec:results}Results and discussion}
\subsection{\label{subsec:results1}Our fit}
\begin{figure}[h]
\centering
\includegraphics[scale=0.91]{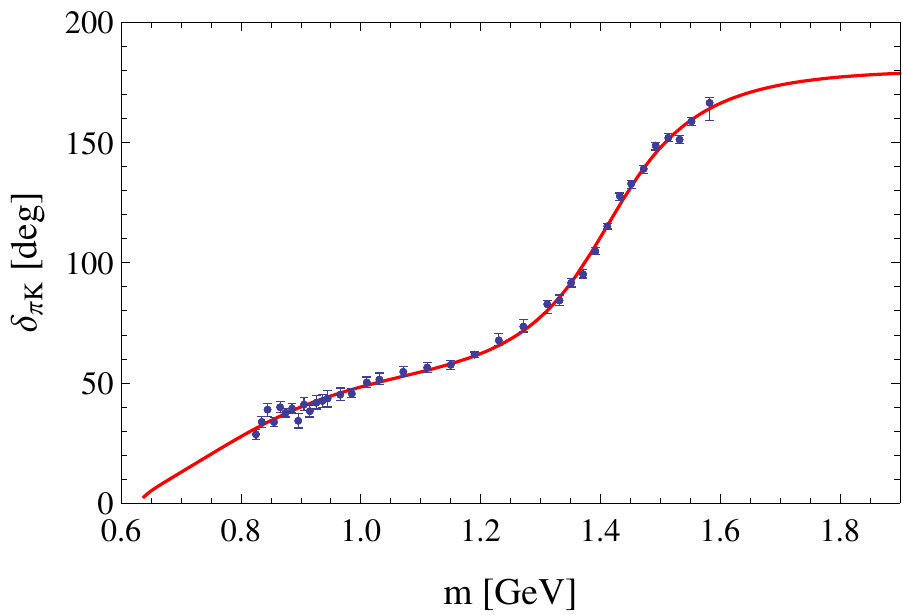}\caption{\label{fig:fit}The solid (red) curve shows our fit
result for the phase shift from Eq.\ (\ref{eq:phaseshift}) with respect to the
four model parameters $a,b,\Lambda,m_{0}$ (see Table\ \ref{tab:parameters}).
The blue points are the data of Ref.\ \cite{pionkaonexp}. A very good agreement
is obtained.}
\end{figure}

The expression from Eq.\ (\ref{eq:phaseshift}) is fitted to the data of Ref.\
\cite{pionkaonexp} with respect to the four model parameters $a,b,\Lambda
,m_{0}$. The result is shown in Fig.\ \ref{fig:fit} and the values of the
parameters together with their errors are reported in Table\
\ref{tab:parameters}. The value of the $\chi^{2}$ is fine: $\chi_{0}
^{2}/d.o.f.=1.25$, explaining the very good agreement of our model result with
data. By comparing the coupling constants it turns out that the derivative
coupling is dominant, which is expected by chPT \cite{chpt,*Ecker,*chpt2} and
by other studies \cite{blackphi,*pagliaraderivatives}.

By using the parameters listed in Table\ \ref{tab:parameters} we continue the
propagator from Eq.\ (\ref{eq:propagator}) into the second Riemann sheet and
scan the complex plane for poles. We find \textit{two} poles (given in GeV)
which we assign in the following way:
\begin{figure*}[t]
\hspace*{-0.75cm}
\begin{minipage}[hbt]{8cm}
\centering
\includegraphics[scale=0.5]{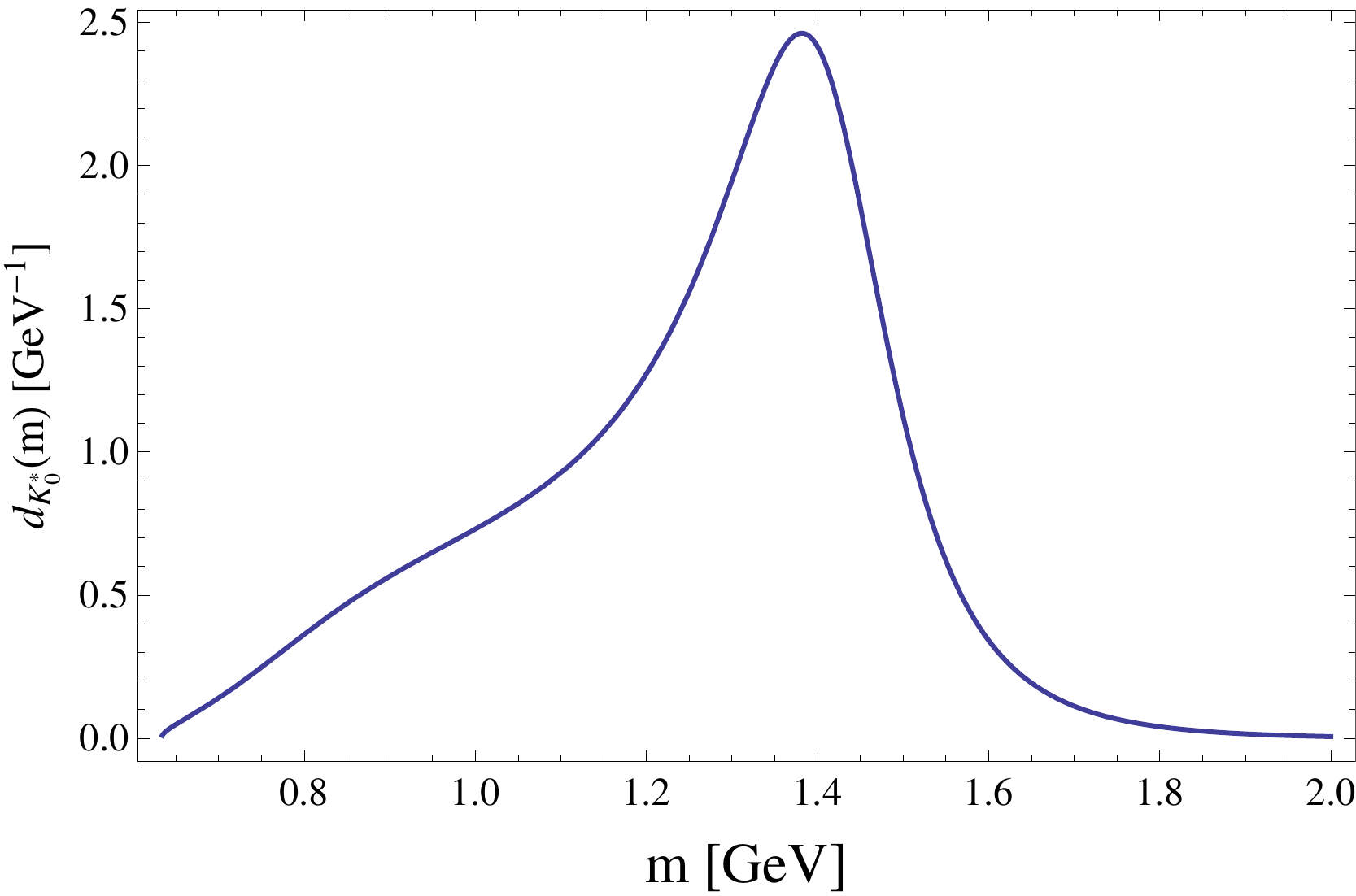}
\end{minipage}
\hspace*{1.0cm}
\begin{minipage}[hbt]{8cm}
\centering
\vspace*{0.11cm}
\includegraphics[scale=0.532]{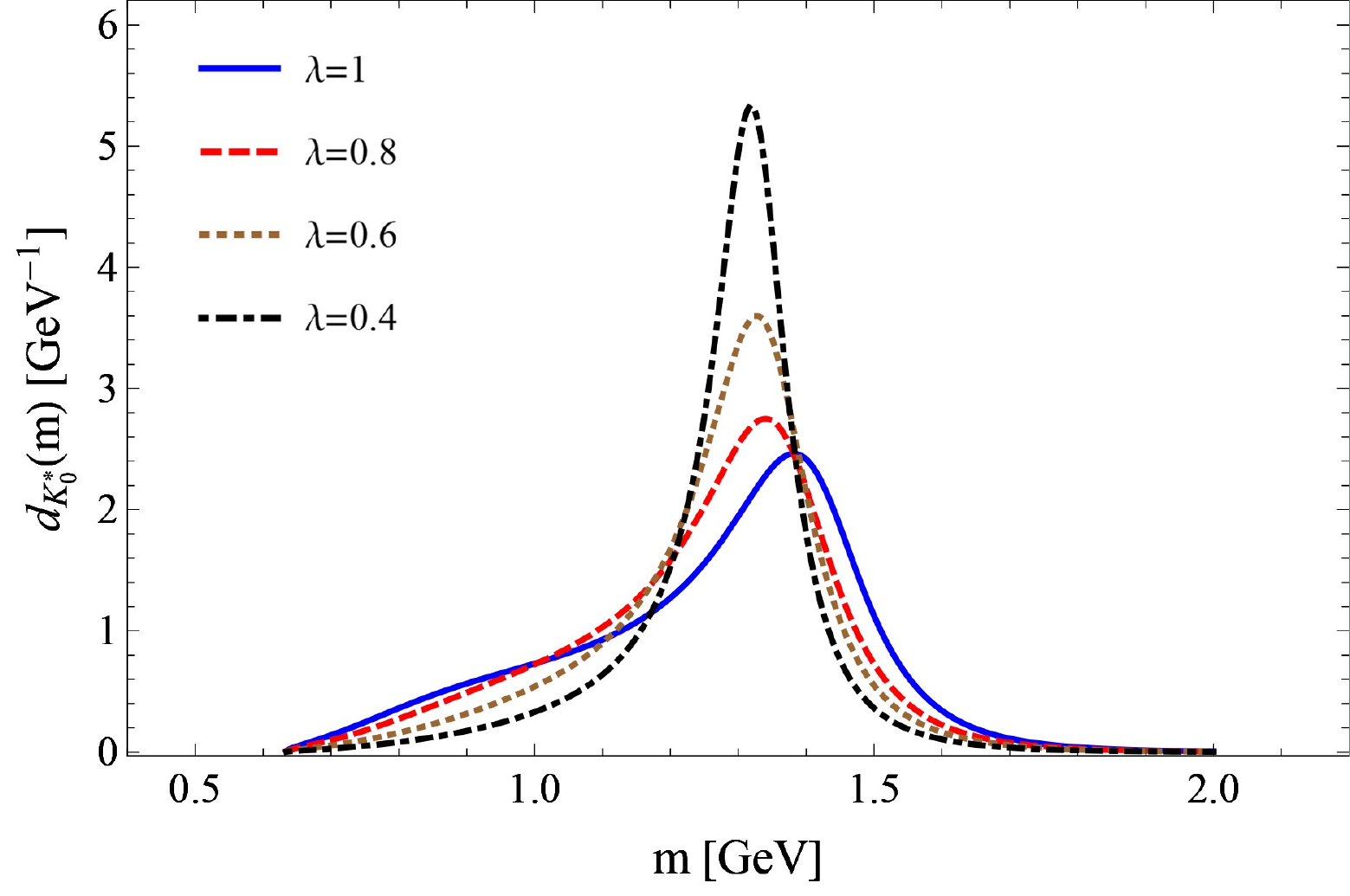}
\end{minipage}
\caption{\label{fig:sf}In the left panel we show the spectral function from Eq.\
(\ref{eq:spectral}) plotted for the parameters of Table\ \ref{tab:parameters}.
An enhancement for low values of the running mass $m$ is clearly visible. In
the right panel we show the spectral function for different values of the
scaling parameter $\lambda=3/N_{c}$. The smaller $\lambda$ is, the more peaked
is the spectral function and the enhancement for low values of the running
mass decreases.}
\end{figure*}
\begin{align}
K_{0}^{\ast}(1430)  & :\ (1.413\pm0.002)-i\hspace{0.02cm}(0.127\pm0.003)\ ,\\
K_{0}^{\ast}(800)\   & :\ (0.746\pm0.019)-i\hspace{0.02cm}(0.262\pm0.014)\ .\label{eq:poles}%
\end{align}
Thus, a pole corresponding to the light $\kappa$ emerges very naturally in our
calculation and is a dynamically generated state (for a discussion on the
definition of dynamical generation, see Refs.\
\cite{giacosaDynamical,guo,ollersigma}). At this point, one should stress that the small errors quoted above (especially for what concerns the resonance $K_{0}^{\ast}(1430)$) are specific to our model defined in Eqs.\ (\ref{eq:lag}),\ (\ref{eq:width}), and\ (\ref{eq:formfactor}), respectively. In particular, the choice of the form factor\ (\ref{eq:formfactor}) is model dependent, a fact that introduces an intrinsic uncertainty. We will explore this point in more detail in the next subsection, in which the positions of the poles are studied for different modifications of the model.
\begin{table}[t]
\caption{\label{tab:parameters}Results of the fit. $\chi_{0}^{2}/d.o.f.=1.25$}%
\begin{ruledtabular}
\begin{tabular}{cl}
Parameter & Value\\
\hline\vspace{-0.25cm}\\
$a$ & $1.60\pm0.22$ GeV\\
$b$ & $-11.16\pm0.82$ GeV$^{-1}$\\
$\Lambda$ & $0.496\pm0.008$ GeV\\
$m_{0}$ & $1.204\pm0.008$ GeV\\
\end{tabular}
\end{ruledtabular}
\end{table}

The PDG \cite{olive} reports for $K_{0}^{\ast}(1430)$ a mass of
$(1.425\pm0.050)$ GeV and a width of $(0.270\pm0.080)$ GeV. Our values fit very
well in these windows. In particular, our width, obtained by doubling the negative 
imaginary part of our pole, reads $(0.254\pm0.006)$ GeV and is thus determined with a small
error. For $K_{0}^{\ast}(800)$ the PDG reports a mass of $(0.682\pm0.029)$ GeV and
a width of $(0.547\pm0.024)$ GeV, which are also in agreement with our values
(although our results point to a somewhat larger value for the mass). The mass 
$(0.746\pm0.019)$ GeV and width 
$(0.524\pm0.028)$ GeV determined within our model are also 
in good agreement with most of the pole determinations listed in Ref.\ \cite{olive}.

In the left panel of Fig.\ \ref{fig:sf} we show the spectral function for the
parameters of Table\ \ref{tab:parameters}. A low-energy enhancement is present,
but no peak. The absence of a peak is one of the reasons why the acceptance of
the $\kappa$ might be considered to be controversial. However, if resonance
poles on unphysical Riemann sheets are the relevant quantities, it turns out
that the existence of the broad $\kappa$ is a consequence of our model.

Similar statements can be made concerning the broad isoscalar state
$f_{0}(500)$: its pole is widely accepted while a clear peak in the spectral
function is not present. On the contrary, the two scalar states $a_{0}(980)$
and $f_{0}(980)$ are pretty narrow: although their couplings are large, these
resonances sit just at the kaon-kaon threshold, making their decays into kaons
to be kinematically suppressed. In conclusion, all those states together with
$\kappa$ seem to have their common origin in quantum fluctuations.

We also study the change of the spectral function and of the position of the
poles when performing a rescaling of the coupling constants:
\begin{equation}
a\rightarrow\sqrt{\lambda}\hspace{0.02cm}a, \ \ b\rightarrow\sqrt{\lambda
}\hspace{0.02cm}b \ \ \ \text{with} \ \lambda\leq1 \ .
\end{equation}
This is completely equivalent to a large-$N_{c}$ study upon setting
\begin{equation}
\lambda= \frac{3}{N_{c}} \ .
\end{equation}
The spectral function is plotted in the right panel of Fig.\ \ref{fig:sf} for
different values of $\lambda$. Obviously, the low-energy enhancement becomes
smaller for decreasing $\lambda$, $i.e.$, for increasing $N_{c}$.

\begin{figure}[b]
\centering
\includegraphics[scale=0.52]{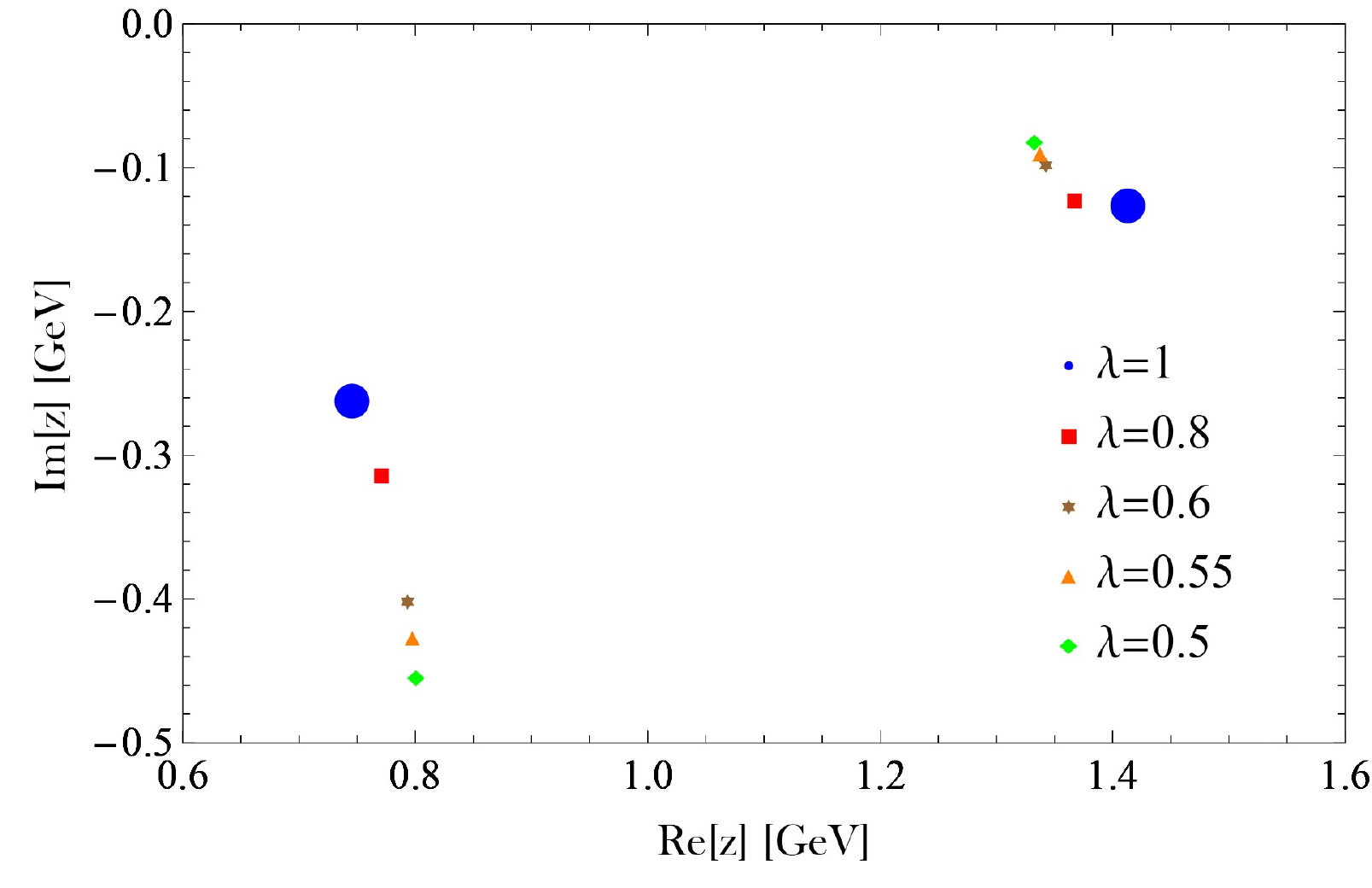}\caption{\label{fig:poles}Movement of the two resonance
poles for different values of $\lambda=3/N_{c}$. While the pole corresponding
to $K_{0}^{\ast}(1430)$ moves toward the real axis, the pole of the light
$K_{0}^{\ast}(800)$ moves away from the real axis and disappears for
$\lambda\simeq0.24$.}%
\end{figure}

Finally, we present the pole movement as function of $\lambda$ in Fig.\
\ref{fig:poles}. We observe that the pole of $K_{0}^{\ast}(1430)$ moves
toward the real axis, a behavior expected for a quarkonium state.
The pole of $K_{0}^{\ast}(800)$ moves away from the real axis and disappears for
$\lambda\simeq0.24$ (or $N_{c}\simeq13$). From this it follows that the pole of
$K_{0}^{\ast}(800)$ is dynamically generated and does not survive in the
large-$N_{c}$ limit. Such a behavior was also reported in Refs.\ \cite{zhiyong,ollerNC,ollerNC2,pelaez3}.

It should be stressed at this point that the choice of the form factor (\ref{eq:formfactor}) is
model dependent. A Gaussian form as implemented here is a standard choice when
investigating mesonic resonances and the position of their poles, respectively,
see also the discussion in Refs.\ \cite{isgurcomment,*tornreply1,harada,*harada2,ruppcomment}. Yet, in 
Sec.\ \ref{subsec:results2} we
investigate possible variations of the form factor and indeed find that they
are not capable of reproducing the phase shift data correctly. At the same
time, we will also investigate the statistical significance of the fit
presented in this subsection as well as the fits that we will discuss in Sec.\ \ref{subsec:results2}.

\subsection{\label{subsec:results2}Variations of the model}
In this subsection we investigate different scenarios in order to understand
better how the results discussed in the previous part emerge. We first perform
two fits to the phase shift data: one in which we consider only the
non-derivative term in Eq.\ (\ref{eq:lag}) (we set $b=0$), and one in which we consider
only the derivative term (we set $a=0$). 

\begin{figure*}[ht!]
\hspace*{-0.95cm}
\begin{minipage}[hbt]{8cm}
\centering
\includegraphics[scale=0.86]{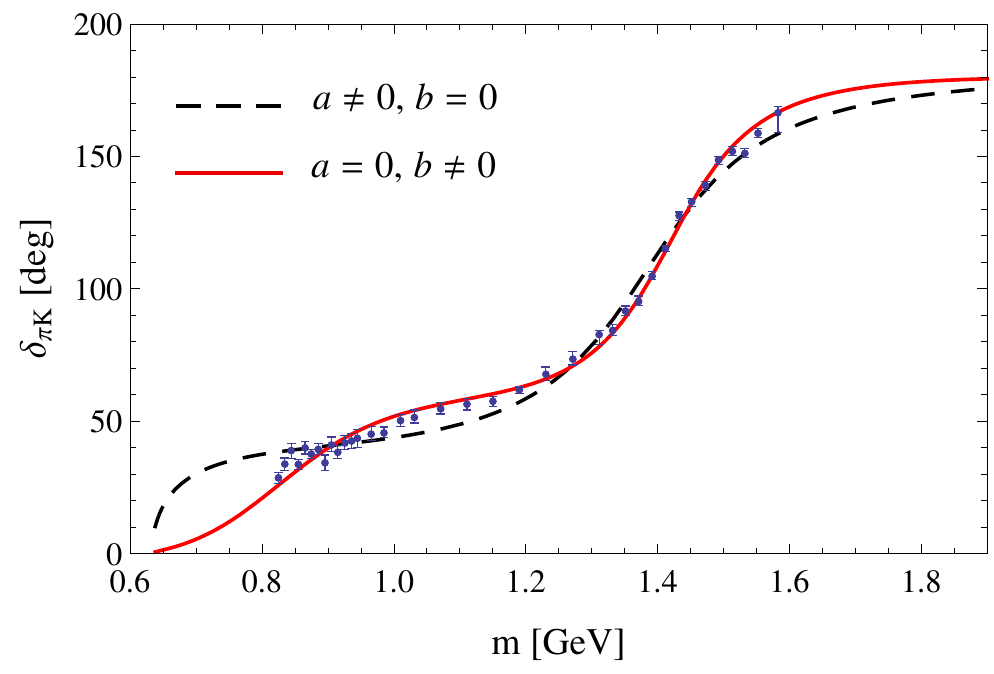}
\end{minipage}
\hspace*{1.0cm}
\begin{minipage}[hbt]{8cm}
\centering
\includegraphics[scale=0.86]{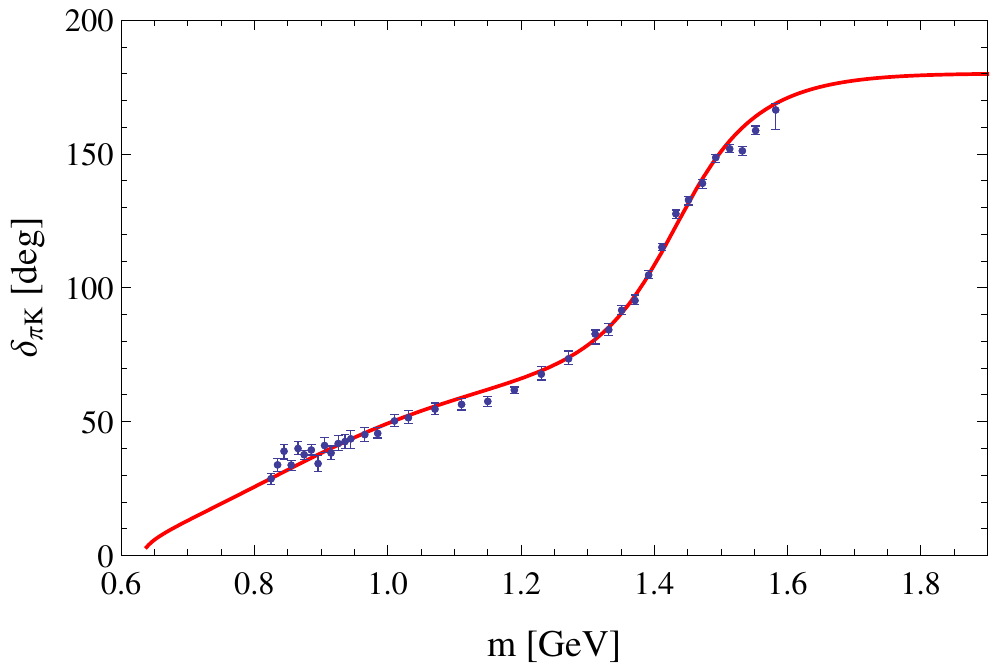}
\end{minipage}
\caption{\label{fig:phases}Left panel: The solid (red) and dashed (black) curves show our fit
results for the phase shift from Eq.\ (\ref{eq:phaseshift}) with respect to the
four model parameters $a,b,\Lambda,m_{0}$ (see Table\ \ref{tab:variations}).
The blue points are the usual data of Ref.\ \cite{pionkaonexp}. Right panel: The solid (red) curve shows the fit for the modified form factor in Eq.\ (\ref{eq:modform}).}
\end{figure*}
The results are presented in the left panel of Fig.\ \ref{fig:phases} and in Table\ \ref{tab:variations}. The first entry
summarizes what was found in the previous subsection. 
The second and third
entries represent the two cases $b=0$ and $a=0$, respectively. As can be seen in the third column, 
in both cases the $\chi^{2}$ has increased, signalizing a worse agreement than with our first fit.
\begin{table*}
\caption{\label{tab:variations}Fitting results for the variations of the model.}
\begin{ruledtabular}
\begin{tabular}{clcccc}
Scenario & \hspace{0.9cm}Parameters & $\chi_{0}^{2}/d.o.f.$ & $p\left(\chi^{2}>\chi_{0}^{2}\right)$ & Pole for $K_{0}^{\ast}(800)$ & Pole for $K_{0}^{\ast}(1430)$\vspace{0.05cm}\\\hline\vspace{-0.25cm}\\
$a,b\neq0$, Gaussian & {$\!\begin{aligned}a &= 1.60\pm0.22 \text{ GeV} \\b &= -11.16\pm0.82 \text{ GeV}^{-1} \\ \Lambda &= 0.496\pm0.008 \text{ GeV} \\m_{0} &=1.204\pm0.008 \text{ GeV} \end{aligned}$} & 1.25 & $0.15$ & {$\!\begin{aligned}(0.746\pm0.019) \\ -i\hspace{0.02cm}(0.262\pm0.014) \end{aligned}$} & {$\!\begin{aligned}(1.413\pm0.002) \\ -i\hspace{0.02cm}(0.127\pm0.003) \end{aligned}$}\vspace{0.05cm}\\\hline\vspace{-0.25cm}\\
$b=0$, Gaussian & {$\!\begin{aligned}a &= 4.06\pm0.04 \text{ GeV} \\ \Lambda &= 0.902\pm0.015 \text{ GeV} \\m_{0} &=1.299\pm0.002 \text{ GeV} \end{aligned}$} & 5.41 & $1.72\cdot10^{-22}$ & - & {$\!\begin{aligned}(1.385\pm0.002) \\ -i\hspace{0.02cm}(0.146\pm0.003) \end{aligned}$}\vspace{0.05cm}\\\hline\vspace{-0.25cm}\\
$a=0$, Gaussian & {$\!\begin{aligned}b &= -17.10\pm0.17 \text{ GeV}^{-1} \\ \Lambda &= 0.453\pm0.002 \text{ GeV} \\m_{0} &=1.142\pm0.002 \text{ GeV} \end{aligned}$} & 2.54 & $1.92\cdot10^{-6}$ & {$\!\begin{aligned}(0.820\pm0.003) \\ -i\hspace{0.02cm}(0.187\pm0.002) \end{aligned}$} & {$\!\begin{aligned}(1.419\pm0.001) \\ -i\hspace{0.02cm}(0.112\pm0.002) \end{aligned}$}\vspace{0.05cm}\\\hline\vspace{-0.25cm}\\
$a,b\neq0$, $F_{\Lambda}(m)=e^{-2k^{4}(m)/\Lambda^{4}}$ & {$\!\begin{aligned}a &= 2.32\pm0.09 \text{ GeV} \\ b &= -3.40\pm0.26 \text{ GeV}^{-1}\\ \Lambda &= 0.652\pm0.006 \text{ GeV} \\m_{0} &=1.248\pm0.003 \text{GeV} \end{aligned}$} & 2.86 & $7.98\cdot10^{-8}$ & {$\!\begin{aligned}(0.863\pm0.008) \\ -i\hspace{0.02cm}(0.339\pm0.017) \end{aligned}$} & {$\!\begin{aligned}(1.433\pm0.002) \\ -i\hspace{0.02cm}(0.112\pm0.003) \end{aligned}$} \\
\end{tabular}
\end{ruledtabular}
\end{table*}

Yet, in order to be more quantitative, we report in the fourth column the
results of a statistical test of the goodness of the fit: The quantity
\begin{equation}
p\left(\chi^{2}>\chi_{0}^{2}\right)=\frac{1}{2^{d/2}\Gamma(d/2)}\int_{\chi_{0}^{2}}^{\infty}\text{d}x \ x^{\frac{d}{2}-1}e^{-x/2}
\end{equation}
(with $d=d.o.f.$) is the probability to obtain a larger value of the $\chi
^{2}$ than $\chi_{0}^{2}$ if a new experiment shall be performed (by using, of course, the same
theoretical function in the fit). When this probability is very small, one may
conclude that $(i)$ the theoretical model is not correct (a reasonable
conclusion) or $(ii)$ the theoretical model is correct but the experimental
results show a -- quite unlucky -- statistical fluctuation. When this probability
is, for instance, smaller than $5\%$, one can exclude the theoretical model at
the $95\%$ confidence level. In our case, our preferred solution from the
previous subsection gives $p\left(\chi^{2}>\chi_{0}^{2}\right)=15.3\%$, which implies that the
theoretical model \textit{cannot} be rejected (here, $d.o.f.=37-4=33$). On
the contrary, the models with only non-derivative interactions and with
derivative interactions can be rejected with a very high level of accuracy.
While this result is expected for the non-derivative term because the shape of
the theoretical function does not match the data (see left panel of Fig.\ \ref{fig:phases}), the situation
is more subtle in the case of only derivative terms. Here, the form is by-eye
qualitatively correct, but the statistical test shows that it is not in
agreement with the experiment (with $d.o.f.=37-3=34$).

Finally, in the fifth and sixth columns we report for completeness the
position of the poles for the various models. Yet, in view of the statistical
analysis, only the first row can be regarded as reliable.

As a next step we investigate other types of the form factor. As explained
before, the Gaussian form factor is rather standard in various works on the
subject and it is also easy to use. Especially in presence of derivative
interactions it is very practical since it cuts off the integrand in the loop
integral sufficiently fast \footnote{When using a form like
$\left(1+k^{n}/\Lambda^{n}\right)^{-2}$ one usually observes non-physical
bumps at high energies\ \cite{tqmix2,*tqmix}.}. However, there is no fundamental reason why the
Gaussian should be the best one to apply. It is therefore important to check
variations of it. We test the following simple modification:
\begin{equation}
F_{\Lambda}(m)=e^{-2k^{4}(m)/\Lambda^{4}}\ .
\label{eq:modform}
\end{equation}
The result of the fit is reported in the right panel of Fig.\ \ref{fig:phases} as well as in the last entry of
Table\ \ref{tab:variations}. Also in this case, the right panel of Fig.\ \ref{fig:phases} shows a qualitative agreement of the model
with data. Yet, the statistical test excludes this model at a very high-level
of accuracy. From this perspective it is not surprising to find the pole of
the $\kappa$ to be not in agreement with our result in the previous subsection
and with other listings in the PDG. Thus, changing the form factor does not
guarantee a good description of data, especially for what concerns the
$\kappa$.

We have also tried a Fermi function $F_{\Lambda
}(m)=\big[(1+e^{-\alpha\Lambda^{2}})/(1+e^{\alpha(k^{2}(m)-\Lambda^{2}%
)})\big]^{2}$ for various values of the parameter $\alpha.\ $This form factor
is approximately constant for small $k$ and rapidly decreases to zero for
$k\sim\Lambda$ (the higher $\alpha$, the steeper the descent; for
$\alpha\rightarrow\infty$ the Heaviside step-function is realized). But also
for this choice it was not possible to obtain a fit which would pass the
statistical test of the $\chi^{2}$.

In conclusion, our study confirms that the Gaussian form factor is an adequate
choice for mesonic interactions, leading to results that are in a good
agreement with the data up to $\sim1.8$ GeV, when \textit{both} a (dominant)
derivative and a (subdominant) non-derivative interaction term are \textit{simultaneously} taken into account.

\section{\label{sec:conclusions}Conclusions}
The scalar sector of hadron physics has been in the center of debate both from
the theoretical and experimental side since a long time. There seems to be a
consensus nowadays that at least the scalar states below $1$ GeV are
non-conventional mesons \cite{amslerrev,*amslerrev2,giacosaDynamical}. In
particular, the role of hadronic loop contributions to the self-energy, such
as the one in Fig.\ \ref{fig:loop}, has been found to be crucial in various
studies
\cite{tornclose,pelaez,*pelaez2,pelaez3,oller,*oller2,*oller3,oller4,oller5,*oller6,2006beveren,morgan,
dullemond,tornqvist,*tornqvist2,boglione,*pennington,giacosapagliara,*e38}.

We have concentrated in this work on the channel $I=1/2$, $J=0$. Our model
contains non-derivative and derivative interactions in agreement with
effective approaches of low-energy QCD \cite{chpt,*Ecker,*chpt2,elsm2}. It was
demonstrated that, by using a single kaonic seed state, both scalar resonances
$K_{0}^{\ast}(1430)$ and $K_{0}^{\ast}(800)$ (known as $\kappa$) can be
described as complex propagator poles. The two poles are required in order to
correctly reproduce phase shift data of $\pi K$ scattering. The spectral
function of our model turns out to be not of the ordinary Breit--Wigner type,
too, due to strong distortions in the low-energy regime, which are a direct
consequence of the $\kappa$-pole.

In the large-$N_{c}$ limit this pole finally disappears; the corresponding
state is therefore not a conventional quarkonium. On the contrary, the pole
corresponding to $K_{0}^{\ast}(1430)$ approaches the real energy axis for
large values of $N_{c}$, hence becomes very narrow, which is a general feature of a
quark-antiquark state.

It must be stressed that the presence of derivative interactions is crucial
for our results. They turn out to be the dominant contribution toward the
description of the $\pi K$ phase shift. For the future, one should use more complete models than the one presented in
this work. In particular, a model is desired which allows to study
simultaneously the $I=1/2$ and the $I=3/2$ channels. For instance, the
extended Linear Sigma Model of Ref.\ \cite{elsm2}, that was used here as a motivation
for our Lagrangian with derivative and non-derivative terms, can be applied
for this purpose. Preliminary results in this direction are encouraging: In
the $I=1/2$ sector this more complete hadronic model reduces -- also for what
concerns the numerical values -- to the Lagrangian of Eq.\ (\ref{eq:lag}).

\begin{acknowledgements}
T.W.\ acknowledges financial support from HGS-HIRe, F\&E GSI/GU, and HIC for FAIR Frankfurt.
\end{acknowledgements}

\bibliography{sk11may}

\end{document}